\documentclass[11pt,preprint,showpacs,preprintnumbers,amsmath,amssymb]{revtex4}

\usepackage{graphicx}
\usepackage{graphics}
\usepackage{dcolumn}
\usepackage{bm}
\usepackage{afterpage}
\usepackage{subfigure}

\usepackage{setspace}
\begin{document}
\title{Supergravity on an Atiyah-Hitchin Base}

\author{Sean Stotyn}
\email{smastoty@sciborg.uwaterloo.ca}
\author{R.B. Mann}
\email{rbmann@sciborg.uwaterloo.ca}

\affiliation{Department of Physics and Astronomy, University of Waterloo \\
             Waterloo, Ontario, Canada, N2L 3G1}

\renewcommand\thesection      {\arabic{section}}
\renewcommand\thesubsection   {\thesection.\arabic{subsection}}
\renewcommand\thesubsubsection  {\thesubsection.\arabic{subsubsection}}
\renewcommand\theequation {\thesection.\arabic{equation}}

\begin{abstract}

We construct solutions to five dimensional minimal supergravity using an Atiyah-Hitchin base space.  In examining the structure of solutions we show that they generically contain a singularity either on the Atiyah-Hitchin bolt or at larger radius where there is a singular solitonic boundary.  However for most points in parameter space the solution exhibits a velocity of light surface (analogous to what appears in a G\"{o}del space-time) that shields the singularity.  For these solutions, all closed time-like curves are causally disconnected from the rest of the space-time in that they exist within the velocity of light surface, which null geodesics are unable to cross.  The singularities in these solutions are thus found to be hidden behind the velocity of light surface and so are not naked despite the lack of an event horizon.  Outside of this surface the space-time is geodesically complete, asymptotically flat and can be arranged so as not to contain closed time-like curves at infinity.  The rest of parameter space simply yields solutions with naked singularities.

\end{abstract}

\maketitle

\section{Introduction}
\label{sec:intro}

Higher dimensional gravity solutions continue to receive a lot of attention because such solutions generally exhibit much richer structure and deeper physics than their four dimensional counterparts.   Amongst their most notable features is their topology.  Black holes in four dimensional space-time can only have a horizon topology of $S^2$, enforced by the black hole uniqueness theorem, whereas in five dimensions such a uniqueness theorem does not exist and black holes can have a horizon topology of either $S^3$, such as the Myers-Perry black hole~\cite{ref:MyersPerry}, or $S^2\times S^1$, such as the black ring found by Emparan and Reall \cite{ref:EmparanReall}.  An example of richer structure in higher dimensions is the black saturn configuration found by Elvang and Figueras  \cite{ref:ElvangFigueras}, which describes a black hole surrounded by a black ring; the rotations of both objects can be chosen so that the total angular momentum is zero, leading to a solution with the same mass and angular momentum as a five dimensional Schwarzschild solution.  In even higher dimensional space-time the horizon topology can become increasingly more complicated and the solutions much more physically rich.

These strictly higher dimensional solutions are important for deepening our understanding of gravitational physics but they are not necessarily the low-energy supersymmetry-preserving states of string theory that we can use to further investigate issues in quantum gravity.  However, coupling gravity to supersymmetry gives us supergravity solutions that can then be embedded in ten or eleven dimensional supergravity theories which in turn may provide consistent backgrounds for string theory.  In Ref.~\cite{ref:Gauntlett}, Gauntlett et al. set out a prescription for generating solutions to five dimensional minimal supergravity.  The solutions fall into two classes depending on whether the Killing vector constructed from the Killing spinor is time-like or null.  There has been a large amount of work recently generating solutions, both by the construction laid out by Gauntlett et al. and by other means such as the Bena-Warner algorithm \cite{ref:BenaWarner}.  The BMPV (Breckenridge-Myers-Peet-Vafa) solution of \cite{ref:BMPV} is constructed on a flat base space and describes an asymptotically flat black hole specified by a mass and two equal angular momenta and exhibits a horizon topology of $S^3$.  Elvang et al. constructed an asymptotically flat supersymmetric black ring solution on a flat base space specified by a mass and two independent angular momenta \cite{ref:Elvangetal}.  The supersymmetric G\"{o}del solution in Ref.\cite{ref:Gauntlett} is also constructed on a flat base space and is the five dimensional supersymmetric analogue of the G\"{o}del universe.  Supersymmetric Kerr- and Schwarzschild-G\"{o}del black holes have been constructed in Ref. \cite{ref:GimonHashimoto}.  In \cite{ref:Ishihara}, Tomizawa et al. used the Eguchi-Hanson space to construct a supersymmetric black ring solution specified by a mass and two equal angular momenta.  In \cite{ref:GauntlettGutowskiI,ref:GauntlettGutowskiII} Gauntlett and Gutowski constructed supersymmetric analogues to the black saturn configuration; their
  solutions describe concentric black rings with an optional black hole
 at the common center.  In \cite{ref:BenaKraus} Bena and Kraus used the Taub-NUT base space to construct a black ring solution that is specified by three charges and three dipole moments. 
 
In the present paper we use the construction described by Gauntlett et al. in Ref.~\cite{ref:Gauntlett} using the Atiyah-Hitchin metric as our hyper-K\"{a}hler base space.  If the base space used admits a Gibbons-Hawking form then the solution is generated by a series of source functions harmonic on the base.  Unlike the other four dimensional hyper-K\"{a}hler metrics (i.e. the flat, Taub-NUT and Eguchi-Hanson spaces) the Atiyah-Hitchin metric cannot be put into a Gibbons-Hawking form and so we must resort to other means.  For simplicity we choose all the metric functions to only be functions of the radial coordinate and the solution so constructed is generated by two first-order differential equations and one Poisson equation on the base.  Although the Atiyah-Hitchin metric depends on the radial coordinate through elliptic integral functions, remarkably Bena et. al. were able to   solve the supergravity equations
analytically via a judicious choice of radial coordinate \cite{ref:Bena} .  We employ these results along with a different ansatz for the 1-form connection to construct and analyze a new solution.  Based on the properties of this solution we show that for most of the parameter space, our space-time describes a region of closed time-like curves which surrounds either a naked singularity or a singular solitonic boundary and which is causally disconnected from the rest of the space-time where observers live.  There is a set of parameter space of measure zero for which such singularities are not so shielded.

This solution seems to be a new type of causal structure similar to, yet importantly distinct from, G\"{o}del and black hole-G\"{o}del solutions.  In such solutions the closed time-like curves exist outside of a certain region and hence extend to asymptotic infinity.  In our solution the closed time-like curves exist within an impenetrable bounded region, causally disconnecting the pathological region from the rest of the space-time, which includes a flat asymptotic infinity.  It seems, then, that the solution presented here has the potential to lead to a well defined holographic dual description (provided such a description exists), free of pathologies, despite the closed time-like curves present in the bulk.  This last issue is perhaps the most tantalizing aspect of our solution and merits further investigation; a discussion of this is postponed until section \ref{sec:Conclusion}.

This paper begins in section \ref{sec:solutions} with a very brief review of the solution generating technique for five dimensional minimal supergravity outlined by Gauntlett et al. in Ref.~\cite{ref:Gauntlett}.  In section \ref{sec:AtiyahHitchin} we provide an overview of the Atiyah-Hitchin metric as well as a brief discussion of its key features.  We go on in section \ref{sec:SUSY} to solve the equations of five dimensional minimal supergravity using the Atiyah-Hitchin metric as the base space.  In section \ref{sec:SUSY}.1 we give analytic expressions for the solution near the center of the space-time and at asymptotic infinity, as well as provide plots of the solution for the rest of the space-time.  In section \ref{sec:SUSY}.2 we analyze the space-time and discuss the generic properties of our solution.  We conclude in section \ref{sec:Conclusion} with a summary of our solution and a discussion of the various issues raised throughout the paper as well as possible future research directions.

\section{Generating Solutions to 5d Minimal Supergravity}
\label{sec:solutions}

\setcounter{equation}{0}

Here we briefly outline the key aspects of the solution generating technique for five dimensional minimal supergravity in order to make our paper more or less self contained; for a complete description the reader is referred to Ref.~\cite{ref:Gauntlett}.  The bosonic sector of five dimensional minimal supergravity is governed by the same action as Einstein-Maxwell theory with an additional Chern-Simons term
\begin{equation}
{\cal S}=\frac{1}{4\pi G}\int{\left(-\frac{1}{4}R\star 1-\frac{1}{2}{\bf F\wedge}\star{\bf F}-\frac{2}{3\sqrt{3}}{\bf F}\wedge {\bf F}\wedge {\bf A}\right)}
\end{equation}
where $G$ is the five dimensional Newton's constant, $\star$ denotes the Hodge dual operator and ${\bf F}={\bf dA}$ is the field strength.  The solutions to the equations of motion of this action are supersymmetric if they admit a Killing spinor from which a real scalar, $q$, real 1-form, ${\bf V}$, and three complex 2-forms, ${\bf \Phi}^{(ab)}$, are constructed.  The solutions are classified according to whether ${\bf V}$, which is also a Killing vector and satisfies $V_\alpha V^\alpha=-q^2$, is time-like or null.  We shall only focus on the time-like case here.  It is further assumed, without loss of generality, that $q>0$ so as to avoid $V^\alpha$ becoming null; the case of $q<0$ has modifications to what follows but the resulting solution is the same in both cases.

If we introduce coordinates such that $V^{\alpha}\partial_{\alpha}=\partial/\partial t$ then the metric can be written locally as
\begin{equation}
ds^2=-H^{-2}(dt+{\bf \omega})^2+Hh_{mn}dx^mdx^n\label{eq:5dmetric}
\end{equation}
where we introduce the scalar $H=q^{-1}$ for convenience and $\omega$ is a 1-form connection.  The metric $Hh_{mn}$ is obtained by projecting the full metric perpendicular to the orbits of $V^{\alpha}$ and furthermore $h_{mn}$ must be hyper-K\"{a}hler with a positive orientation chosen so that the hyper-K\"ahler structures (related to the  ${\bf \Phi}^{(ab)}$) are anti-self-dual.  We will thus denote $h_{mn}dx^mdx^n=ds_{\cal B}^2$ where the base space, $\cal B$, is endowed with a hyper-K\"{a}hler metric.  

We next define
\begin{equation}
{\bf e^0}={H^{-1}}(dt+\omega)
\end{equation}
so that if $\sigma$ defines the proper positive orientation on $\cal B$ then ${\bf e^0}\wedge\sigma$ will define a positive orientation on the five dimensional space-time.  With this definition, the form for ${\bf F}$ is
\begin{equation}
{\bf F}=\frac{\sqrt{3}}{2}{\bf de^0}-\frac{1}{\sqrt{3}}{\bf G^+}
\end{equation}
where ${\bf G^+}$ is a self dual 2-form on $\cal B$ defined via
\begin{equation}
{\bf G^+}=\frac{H^{-1}}{2}({\bf d\omega}+\star_4 {\bf d\omega}). \label{eq:Gomega}
\end{equation}
Here $\star_4$ denotes the Hodge dual on the four dimensional space $\cal{B}$.  

The Bianchi identity and the equation of motion for {\bf F} respectively give
\begin{eqnarray}
&&{\bf dG^+}=0 \label{eq:Bianchi}\\
&&\Delta H=\frac{4}{9}({\bf G^+})^2 \label{eq:Poisson}
\end{eqnarray}
where 
\begin{equation}
\Delta=\frac{1}{\sqrt{g}}\partial_i(\sqrt{g}g^{ij}\partial_j) \label{eq:Laplace}
\end{equation}
is the Laplacian operator on the base space.  The consequence of Eq.~(\ref{eq:Bianchi}) is that we can write ${\bf G^+=\alpha d\Omega}$, where $\alpha$ is a constant and ${\bf \Omega}$ is a 1-form. 

If the base space admits a  triholomorphic Killing vector field then its metric can be put into Gibbons-Hawking form and the five dimensional supergravity solutions are generated by four arbitrary functions harmonic on the base if ${\bf G^+}=0$ or three if ${\bf G^+}\ne0$.  An example of a hyper-K\"{a}hler base space which admits a triholomorphic Killing vector field is the Eguchi-Hanson space and in Ref.~\cite{ref:Ishihara} Tomizawa et al. constructed a supersymmetric black ring on this space exploiting its Gibbons-Hawking form.  The Atiyah-Hitchin space, however, does not admit a Gibbons-Hawking metric and so we must grind through the machinery of this section to obtain our solution.

\section{The Atiyah-Hitchin Space}
\label{sec:AtiyahHitchin}

\setcounter{equation}{0}

The dynamics of two non-relativistic BPS (Bogomol'nyi-Prasad-Sommerfield) monopoles is described by a manifold $\cal{M}$ which has the product structure
\begin{equation}
{\cal{M}}=\mathbb{R}^3\times\frac{S^1\times M}{\mathbb{Z}_2}
\end{equation}
where a point in $\mathbb{R}^3\times S^1$ denotes the centre of mass of the system and a time-varying phase angle that determines the total electric charge, while a point in $M$ specifies the monopole separation and a relative phase angle.  The four dimensional manifold, $M$, is invariant under SO(3) and can be parameterised by a radial coordinate $r$, roughly giving the separation of the monopoles, and Euler angles $\theta$, $\phi$ and $\psi$ with ranges $0\le\theta\le\pi$, $0\le\phi\le 2\pi$, and $0\le\psi\le 2\pi$.

We introduce a basis for SO(3) via the left invariant Maurer-Cartan 1-forms, which are related to the Euler angles through the relations
\begin{eqnarray}
&&\sigma_1^R=-\sin\psi d\theta +\cos\psi\sin\theta d\phi \nonumber\\
&&\sigma_2^R=\cos\psi d\theta +\sin\psi\sin\theta d\phi \\
&&\sigma_3^R=d\psi +\cos\theta d\phi \nonumber
\end{eqnarray}
and which have the property
\begin{equation}
d\sigma_i^R=\frac{1}{2}\varepsilon_{ijk}\sigma_j^R\wedge\sigma_k^R.
\end{equation}
For convenience, in further discussion we drop the superscript $R$.  In terms of these 1-forms, the flat metric on $\mathbb{R}^4$ is given by $ds^2=dR^2+\frac{R^2}{4}({\sigma_1}^2+{\sigma_2}^2+{\sigma_3}^2)$, the metric on a unit radius $S^3$ is given by $d{\Omega_3}^2=4({\sigma_1}^2+{\sigma_2}^2+{\sigma_3}^2)$ and the metric on a unit radius $S^2$ is given by $d{\Omega_2}^2={\sigma_1}^2+{\sigma_2}^2$. The metric on $M$, known as the Atiyah-Hitchin metric, can be written in the explicitly SO(3) invariant form
\begin{equation}
ds^2=f^2(r)dr^2+a^2(r){\sigma_1}^2+b^2(r){\sigma_2}^2+c^2(r){\sigma_3}^2\label{eq:AH}.
\end{equation}
As shown in \cite{ref:GibbonsPope} the above metric satisfies the vacuum Einstein equations and is self dual, which in four dimensions ensures it is hyper-K\"{a}hler, if $a(r)$, $b(r)$, and $c(r)$ satisfy the differential equations
\begin{equation}
\frac{d}{dr}a(r)=f(r)\frac{(b(r)-c(r))^2-a^2(r)}{2b(r)c(r)}
\end{equation}
plus the two equations obtained by cyclically permuting $a(r)$, $b(r)$, and $c(r)$.  As can be easily verified, if we impose $a(r)=b(r)=c(r)$ then the constraint equations yield the flat metric on $\mathbb{R}^4$ given above.  Similarly, it can easily be checked that choosing any two of $a(r)$, $b(r)$ and $c(r)$ being equal (for example $a(r)=b(r)\ne c(r)$) the resulting metric is the Euclidean Taub-NUT metric.  Imposing the condition that none of the functions are equal yields the Atiyah-Hitchin metric.  

The function $f(r)$ defines the radial coordinate and hence we are able to freely choose its form.  In Ref.~\cite{ref:AtiyahHitchin}, Atiyah and Hitchin used $f(r)=a(r)b(r)c(r)$ to obtain the original form of their solution.  If we instead take $f(r)=-b(r)/r$ then the solution simplifies.  We set
\begin{equation}
r=2nK\left(\sin(\frac{\gamma}{2})\right) \label{eq:r}
\end{equation}
where $K(k)$, and $E(k)$ encountered shortly, are the complete elliptic integrals of the first and second kind respectively
\begin{eqnarray}
&&K(k)=\int_0^{\pi/2}{\frac{dy}{\sqrt{1-k^2\sin^2y}}}\\
&&E(k)=\int_0^{\pi/2}{\sqrt{1-k^2\sin^2y}dy}.
\end{eqnarray}
As $\gamma$ takes value in the range $[0,\pi]$, $r$ takes on values in the range $[n\pi,\infty)$.  If we now define
\begin{eqnarray}
w_1(r)=b(r)c(r), \>\>\>\>\>\>\>\>\>
w_2(r)=c(r)a(r), \>\>\>\>\>\>\>\>\>
w_3(r)=a(r)b(r)\label{eq:w}
\end{eqnarray}
as well as
\begin{equation}
\Upsilon(r)\equiv\frac{dr}{d\gamma}=\frac{2nE\left(\sin(\frac{\gamma}{2})\right)}{\sin(\gamma)}-\frac{nK\left(\sin(\frac{\gamma}{2})\right)\cos(\frac{\gamma}{2})}{\sin(\frac{\gamma}{2})}
\end{equation}
then the solutions for $w_1(r)$, $w_2(r)$ and $w_3(r)$ are given by
\begin{eqnarray}
&&w_1(r)=-r\Upsilon(r)\sin(\gamma)-r^2\cos^2(\frac{\gamma}{2})\nonumber\\
&&w_2(r)=-r\Upsilon(r)\sin(\gamma)\label{eq:wsols}\\
&&w_3(r)=-r\Upsilon(r)\sin(\gamma)+r^2\sin^2(\frac{\gamma}{2}).\nonumber
\end{eqnarray}

\begin{figure}
\centering
\includegraphics[width=4.6 in]{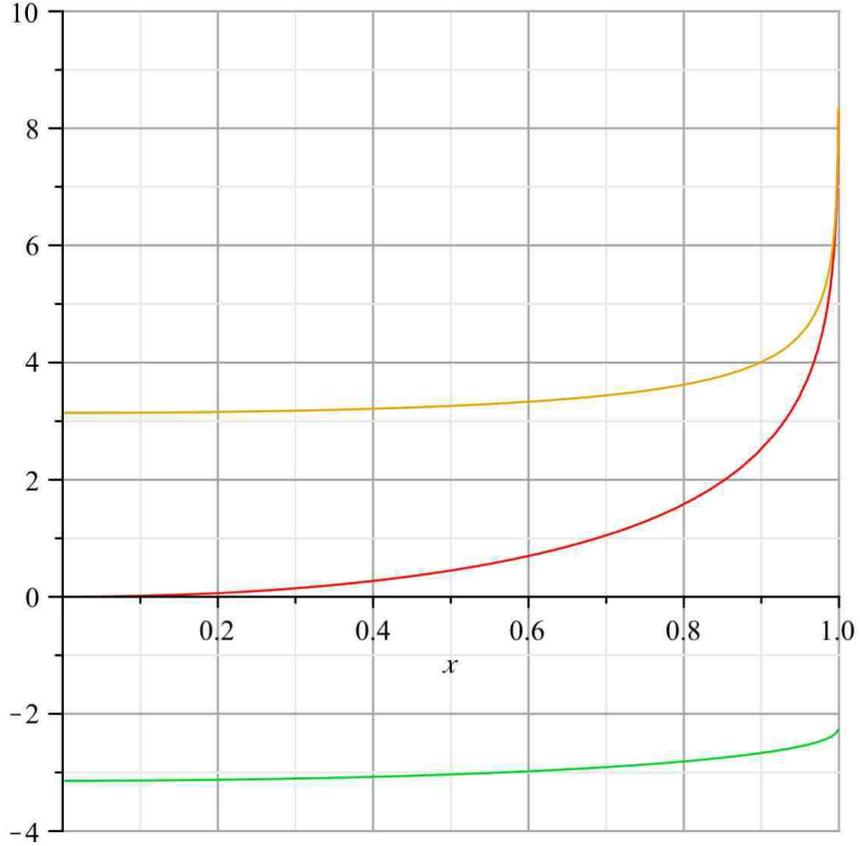}
\caption{The metric functions $a(r)$ (red), $b(r)$ (orange) and $c(r)$ (green) plotted as functions of the variable $x$ defined in Eq.~(\ref{eq:x}).  Notice that $a(r)$ and $b(r)$ asymptote to the same curve for large $r$ (i.e. $x\rightarrow1$) while $c(r)$ approaches a constant.  In the opposite limit of $r\rightarrow n\pi$ (i.e. $x\rightarrow0$) we have $a(r)\rightarrow0$ while $b(r)$ and $c(r)$ approach the same (but opposite sign) constant value.  These behaviours are explicitly shown in Eqs. (\ref{eq:approx}) and (\ref{eq:abclarge}).
\label{fig:abc}}
\end{figure}

We could in principle substitute Eq. (\ref{eq:r})  in the above expressions to get functions only dependent upon $\gamma$ since we cannot invert Eq.~(\ref{eq:r}) to get an explicit solution for $\gamma(r)$.  For numerical computations it is simpler to work in terms of the coordinate 
\begin{equation}
x\equiv \sin\left(\frac{\gamma}{2}\right). \label{eq:x}
\end{equation}
In performing analysis of the structure of space-time (and various physical observables) near the points $r=n\pi$ and $r\rightarrow\infty$ it is easiest to work in terms of $r$ by inverting $\gamma(r)$ via a Taylor expansion.  We thus will not worry about the implicit dependence on $r$ in what follows.  The metric functions $a(r)$, $b(r)$, and $c(r)$, obtained by solving (\ref{eq:w}) and (\ref{eq:wsols}), take the explicit form
\begin{eqnarray}
&&a(r)=\sqrt{\frac{r\Upsilon(r)\sin(\gamma)\left(r\sin^2(\frac{\gamma}{2})-\Upsilon(r)\sin(\gamma)\right)}{r\cos^2(\frac{\gamma}{2})+\Upsilon(r)\sin(\gamma)}}\\
&&b(r)=\sqrt{\frac{\left(r\cos^2(\frac{\gamma}{2})+\Upsilon(r)\sin(\gamma)\right)r\left(r\sin^2(\frac{\gamma}{2})-\Upsilon(r)\sin(\gamma)\right)}{\Upsilon(r)\sin(\gamma)}}\\
&&c(r)=-\sqrt{\frac{r\Upsilon(r)\sin(\gamma)\left(r\cos^2(\frac{\gamma}{2})+\Upsilon(r)\sin(\gamma)\right)}{r\sin^2(\frac{\gamma}{2})-\Upsilon(r)\sin(\gamma)}}.
\end{eqnarray}
Plots of these functions are given in Fig.~\ref{fig:abc}.  

The Taylor expansions $K(k)\approx \frac{\pi}{2}(1+\frac{1}{4}k^2)$ and $E(k)\approx \frac{\pi}{2}(1-\frac{1}{4}k^2)$ valid near $k=0$ can be used to obtain approximate forms for the metric functions near $r=n\pi$
\begin{eqnarray}
&&a(r)= 2(r-n\pi)\left(1-\frac{1}{4n\pi}(r-n\pi)+{\cal O}((r-n\pi)^2)\right)\nonumber\\
&&b(r)= n\pi\left(1+\frac{1}{2n\pi}(r-n\pi)+{\cal O}((r-n\pi)^2)\right)\label{eq:approx}\\
&&c(r)= -n\pi\left(1-\frac{1}{2n\pi}(r-n\pi)+{\cal O}((r-n\pi)^2)\right).\nonumber
\end{eqnarray}
We make note of the fact that $a(r)\rightarrow 0$ as $r\rightarrow n\pi$ and we can get a better picture of what this means if we rotate our axes such that
\begin{eqnarray}
&&\sigma_1=d\tilde\psi +\cos\tilde\theta d\tilde\phi \nonumber\\
&&\sigma_2=-\sin\tilde\psi d\tilde\theta +\cos\tilde\psi\sin\tilde\theta d\tilde\phi \label{eq:rotate}\\
&&\sigma_3=\cos\tilde\psi d\tilde\theta +\sin\tilde\psi\sin\tilde\theta d\tilde\phi\nonumber.
\end{eqnarray}
Using these rotated axes along with the leading order approximations of (\ref{eq:approx}) the metric near $r=n\pi$ becomes
\begin{equation}
ds^2\approx dr^2+4(r-n\pi)^2(d\tilde\psi +\cos\tilde\theta d\tilde\phi)^2+(n\pi)^2(d\tilde\theta^2+\sin^2\tilde\theta d\tilde\phi^2).
\end{equation}
The third term is the metric on a two-sphere of radius $n\pi$ while the second term vanishes at $r=n\pi$.  The three dimensional SO(3) orbit collapses to a two-sphere of radius $n\pi$ at $r=n\pi$ and so the center of the Atiyah-Hitchin metric is a Bolt.  In the monopole picture, the Bolt corresponds to when the monopoles coincide\cite{ref:Gibbons}.

\afterpage{\clearpage} 

The opposite limit (corresponding to large monopole separation) is that of $r\rightarrow\infty$.  In this limit the metric functions $a(r)$, $b(r)$, and $c(r)$ become
\begin{eqnarray}
&&a(r)=r\sqrt{1-2n/r}+{\cal O}(e^{-r/n})\nonumber\\
&&b(r)=r\sqrt{1-2n/r}+{\cal O}(e^{-r/n})\label{eq:abclarge}\\
&&c(r)=-\frac{2n}{\sqrt{1-2n/r}}+{\cal O}(e^{-r/n})\nonumber
\end{eqnarray}
and the metric (\ref{eq:AH}) reduces to
\begin{equation}
ds^2\approx\left(1-\frac{2n}{r}\right)(dr^2+r^2d\theta^2+r^2\sin^2\theta d\phi^2)+4n^2\left(1-\frac{2n}{r}\right)^{-1}(d\psi+\cos\theta d\phi)^2.
\end{equation}
Note that this is none other than the Euclidean Taub-NUT metric with NUT charge $N=-n$; asymptotically the Atiyah-Hitchin metric describes a Taub-NUT space with negative NUT charge.  This is not surprising since at asymptotic infinity we have $a(r)=b(r)$ and we noted previously that if any two of the metric functions are equal the constraint equations determine the resulting metric to be Taub-NUT.

The last items necessary for our analysis are the positive orientation and the volume element of the Atiyah-Hitchin metric.   The orientation of the base space must be such that the hyper-K\"ahler 2-forms are anti-self-dual.  Such a positive orientation for Taub-NUT is given by $Ndr\wedge\sigma_1\wedge\sigma_2\wedge\sigma_3$ where $N$ is the NUT charge\cite{ref:Gauntlett}. Since the Atiyah-Hitchin metric is asymptotically Taub-NUT with negative NUT charge, we establish the vierbein
\begin{eqnarray}
&&{\bf e}^r=-f(r)dr,\>\>\>\>\>\>\>\>
{\bf e}^1=a(r)\sigma_1,\\
&&{\bf e}^2=b(r)\sigma_2,\nonumber\>\>\>\>\>\>\>\>\>\>\>\>
{\bf e}^3=c(r)\sigma_3,
\end{eqnarray}
so that positive orientation is given by the volume 4-form ${\bf vol}(g)={\bf e}^r\wedge {\bf e}^1\wedge {\bf e}^2\wedge {\bf e}^3$.  After expanding in terms of $r, \theta, \phi$ and $\psi$ this becomes ${\bf vol}(g)=f(r)a(r)b(r)c(r)\sin\theta dr\wedge d\theta\wedge d\phi\wedge d\psi$ from which we immediately see that
\begin{equation}
\sqrt{g}=f(r)a(r)b(r)c(r)\sin\theta.
\end{equation}

\section{SUSY Solution with an Atiyah-Hitchin Base} 
\label{sec:SUSY}

\setcounter{equation}{0}

 We are now equipped with the tools necessary to generate five dimensional supergravity solutions on an Atiyah-Hitchin base.  Unlike the other hyper-K\"{a}hler metrics the Atiyah-Hitchin metric is rather complicated in its dependence upon the radial coordinate.   The first subsection here is devoted to finding the differential equations that must be obeyed in order to produce valid solutions and gives the results found by Bena et al. \cite{ref:Bena} in more convenient coordinates for our analysis.  It also contains approximate forms of the solutions to these differential equations in the limits $r\approx n\pi$ and $r\rightarrow\infty$.  We make note of the fact that we are explicitly choosing $H$ to be a function of $r$ only since it satisfies Poisson's equation on the base and is not separable when it is also a function of angle.  More solutions with an Atiyah-Hitchin base space exist, corresponding to other choices of $H$, but such solutions are not studied here.

\subsection{Finding the Equations and Their Asymptotes}
\label{subsec:Asymptotes}

\begin{figure}
\centering
\includegraphics[width=4.0 in]{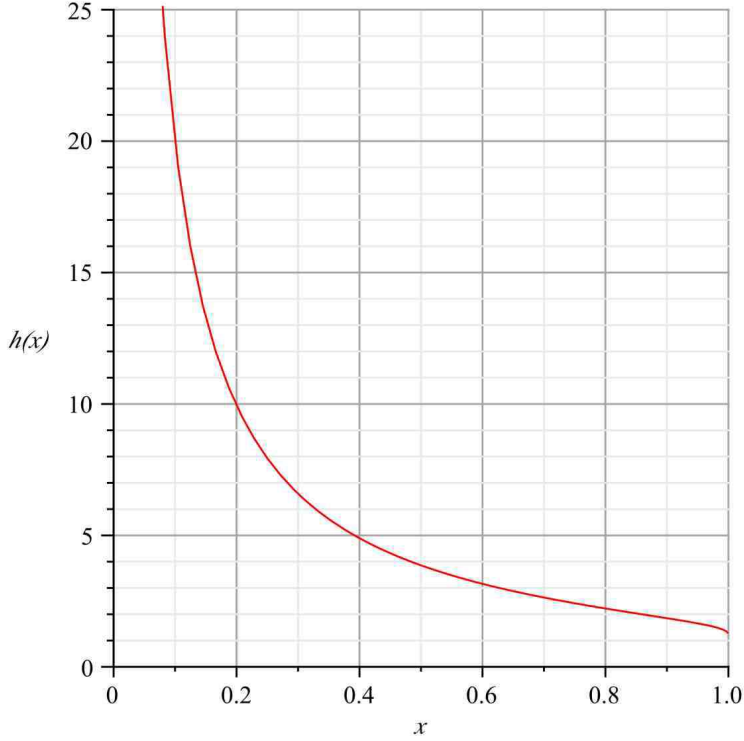}
\caption{The behaviour of h(r) in terms of the variable $x$ defined via Eq. (\ref{eq:x}).
\label{fig:smallh}}
\end{figure}

First we seek a solution for ${\bf G^+=\alpha d\Omega}$ which also satisfies Eq.~(\ref{eq:Gomega}).  We thus have to choose a suitable ansatz for both the 1-forms $\omega$ and ${\bf \Omega}$; following \cite{ref:Gauntlett} we consider
\begin{eqnarray}
&&\omega=\Psi(r)\sigma_3\label{eq:omega}\\
&&{\bf \Omega}=h(r)\sigma_3\label{eq:Omega}
\end{eqnarray}
where $\Psi(r)$ and $h(r)$ are arbitrary functions.
Eq.~(\ref{eq:Omega}) along with ${\bf G^+=\alpha d\Omega}$ gives
\begin{equation}
{\bf G^+}=n\chi_0\left(\frac{-h'(r)}{f(r)c(r)}{\bf e}^r\wedge {\bf e}^3+\frac{h(r)}{a(r)b(r)}{\bf e}^1\wedge {\bf e}^2\right)
\end{equation}
where $\chi_0=\alpha/n$ is a dimensionless constant and a prime denotes differentiation with respect to $r$.  The self-duality of ${\bf G^+}$ requires that we must have $\frac{-h'(r)}{f(r)c(r)}=\frac{h(r)}{a(r)b(r)}$ and hence $h(r)$ is given by
\begin{equation}
h(r)=\exp\left(-\int{\frac{f(r)c(r)dr}{a(r)b(r)}}\right).\label{eq:h}
\end{equation}
In Ref.~\cite{ref:Bena}, Bena et al. choose the radial coordinate, $\eta$, such that $f(\eta)=a(\eta)b(\eta)c(\eta)$, which is related to our radial coordinate, $r$, by $\eta=-\int{\frac{dr}{ra(r)c(r)}}$.  Their analytic solution for $h(\eta)$, re-expressed in terms of $r$, is given by
\begin{equation}
h(r)=\frac{r^2\sin\left(\frac{\gamma}{2}\right)}{a(r)b(r)}\label{eq:h2}
\end{equation}
where we have imposed $h(r\rightarrow\infty)=1$ since any constant factors can be absorbed into $\chi_0$.  A plot of Eq. (\ref{eq:h2}) is plotted in Fig.~\ref{fig:smallh} to demonstrate its behaviour.  We can now write ${\bf G^+}$ explicitly as
\begin{equation}
{\bf G^+}=n\chi_0\left(\frac{h(r)}{a(r)b(r)}\right)({\bf e}^r\wedge {\bf e}^3+{\bf e}^1\wedge {\bf e}^2).\label{eq:Gplus}
\end{equation}

\begin{figure}
     \centering
     \subfigure[parameter choices: $\chi=0.1$, $\mu=\{10$ (red), 5 (orange), 0 (green), -5 (cyan), -10 (blue)\}]{\includegraphics[width=3.2 in]{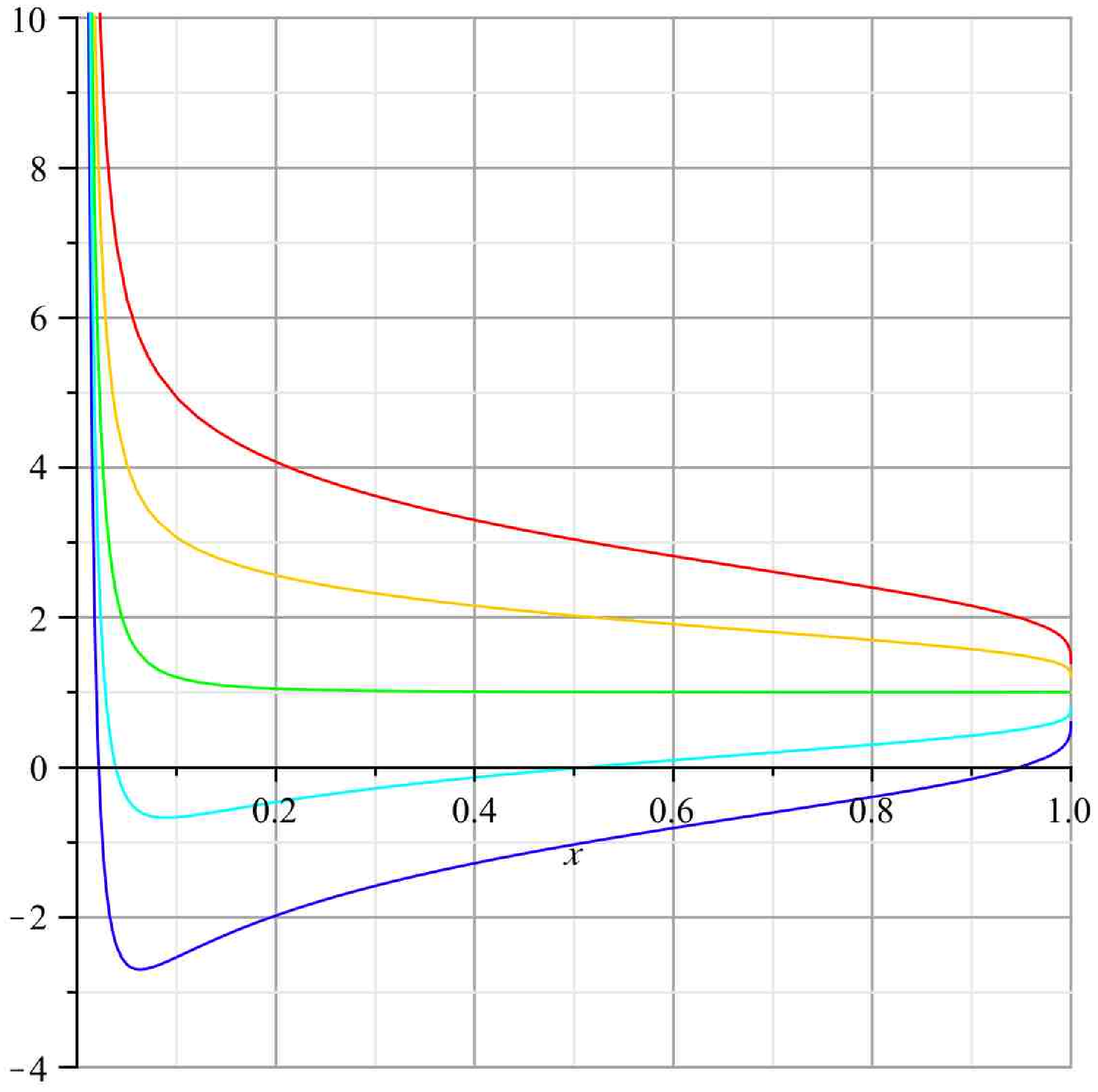}}
     \subfigure[parameter choices: $\mu=0$, $\chi=\{2.5$ (red), 2 (orange), 1.5 (green), 1 (cyan), 0.5 (blue)\}]{\includegraphics[width=3.2 in]{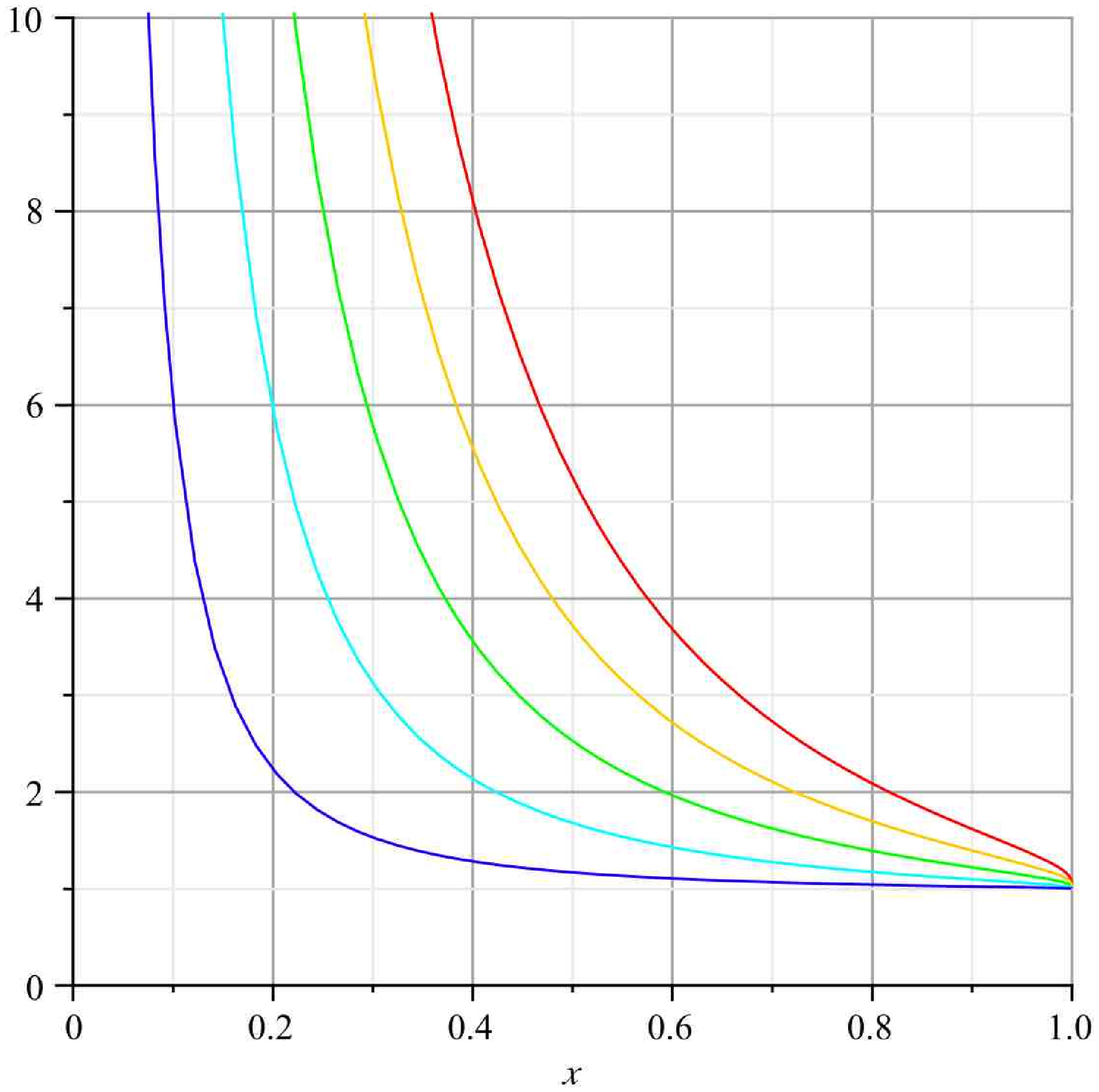}}
     \subfigure[parameter choices: $\chi=0$, $\mu=\{10$ (red), 5 (orange), 0 (green), -5 (cyan), -10 (blue)\}]{\includegraphics[width=3.2 in]{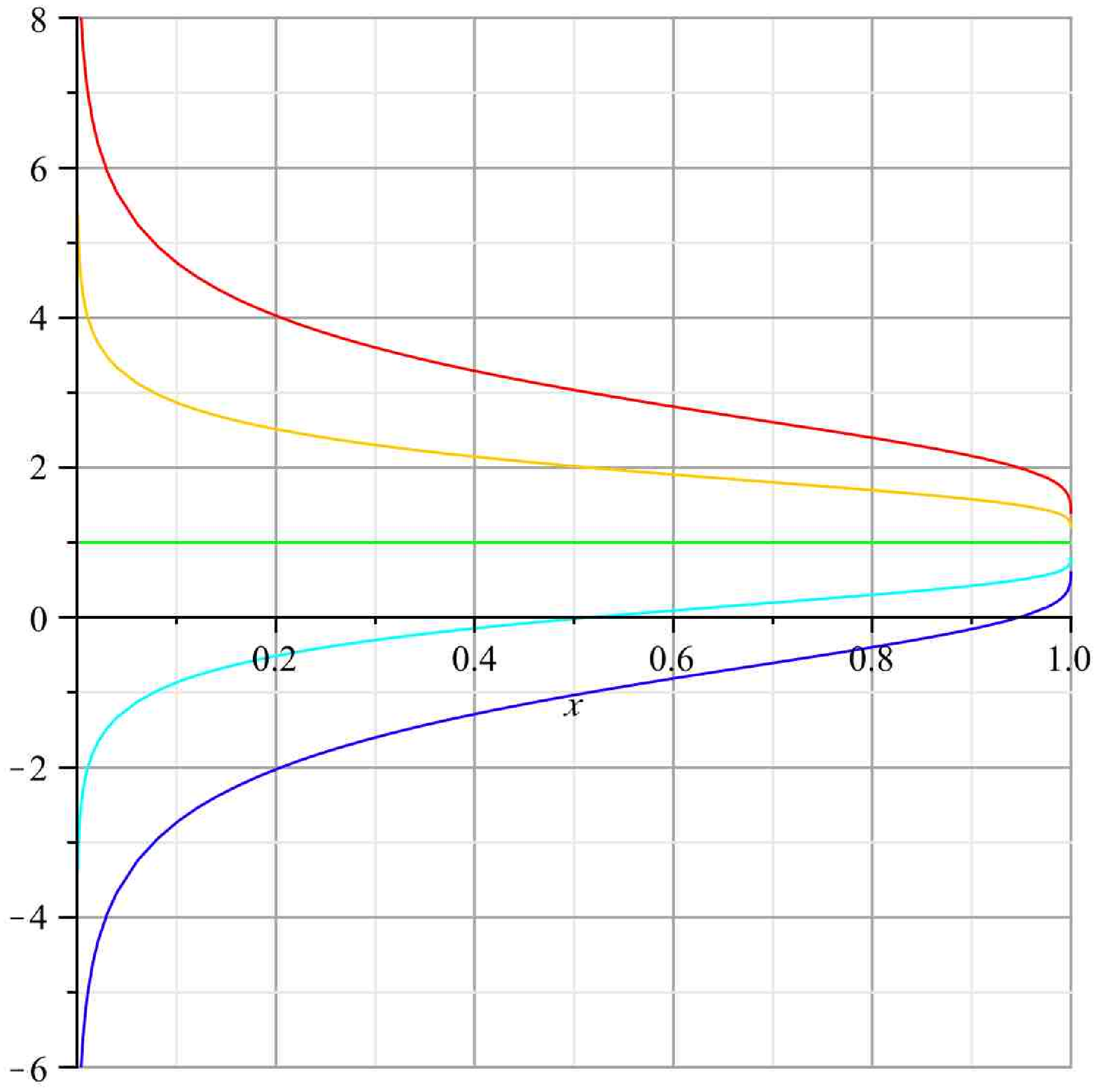}} \\
     \caption{The behaviour of $H(r)$ for the specified choices of parameters.  Note that $H(r)$ can become negative: this point will be interpreted and expanded on later.}
     \label{fig:H}
\end{figure}

With this solution for ${\bf G^+}$ we can use Eq.~(\ref{eq:Poisson}) to find the differential equation for $H(r)$:
\begin{equation}
\frac{1}{f(r)a(r)b(r)c(r)\sin\theta}\partial_i\left(f(r)a(r)b(r)c(r)\sin\theta g^{ij}\partial_jH(r)\right)=\frac{8}{9}n^2{\chi_0}^2\left(\frac{h(r)}{a(r)b(r)}\right)^2.
\end{equation}
Since $H=H(r)$ and $g_{rj}=0$ unless $j=r$ the only non zero terms in the Laplacian are $i=j=r$ and we find
\begin{equation}
\frac{d}{dr}\left(\frac{a(r)b(r)c(r)}{f(r)}\frac{dH(r)}{dr}\right)=\frac{8}{9}n^2{\chi_0}^2\left(h^2(r)\frac{f(r)c(r)}{a(r)b(r)}\right).
\end{equation}
From (\ref{eq:h}) we have $h^2(r)\frac{f(r)c(r)}{a(r)b(r)}=-\frac{1}{2}\frac{d}{dr}(h^2(r))$ so the first integration is trivial.  The second order differential equation thus reduces to a first order equation and the solution is given by
\begin{equation}
H(r)=\delta -n^2\lambda\int{\frac{f(r)dr}{a(r)b(r)c(r)}}-\frac{4}{9}n^2{\chi_0}^2\int{\frac{h^2(r)f(r)dr}{a(r)b(r)c(r)}}
\label{eq:H}
\end{equation}
where $\lambda$ and $\delta$ are dimensionless constants of integration.  In \cite{ref:Bena} it is shown that 
\begin{equation}
\frac{d}{dr}\left(\frac{(8\pi^2)^2}{a(r)b(r)}\right)=\frac{f(r)}{a(r)b(r)c(r)}((8\pi^2)^2-4h^2(r))\label{eq:hsq}
\end{equation}
so the last term in (\ref{eq:H}) can be integrated exactly, giving
\begin{equation}
H(r)=\delta -n^2\mu\int{\frac{f(r)dr}{a(r)b(r)c(r)}}+\frac{n^2\chi^2}{a(r)b(r)}
\end{equation}
where $\mu=\lambda+\chi^2$ and $\chi=\frac{8\pi^2}{3}{\chi_0}$.  The integral above we can recognize as $\eta(r)$ but it cannot be written in an analytic form; in \cite{ref:Bena} $\eta$ is also written in terms of an integral of elliptic functions and there is no obvious way to obtain an analytic expression in terms of known functions.  This term is easy to numerically integrate, however, and plots corresponding to various choices of parameters are given in Fig. \ref{fig:H}.  For ease of notation, we will write $\eta(r)$ instead of the integral representation in what follows.

\afterpage{\clearpage}

\begin{figure}
     \centering
     \subfigure[parameter choices: $\mu=0$, $\chi=0$, $j=\{ 2$ (red), 1 (orange), -1 (green), -2 (blue)\}]{\includegraphics[width=3.2 in]{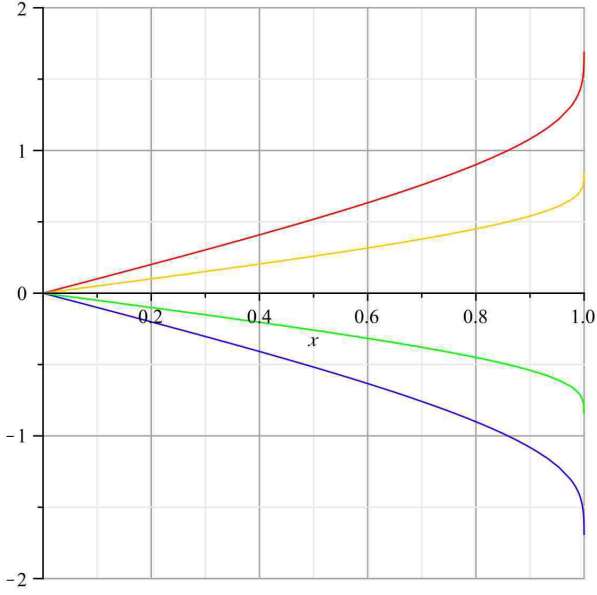}}
     \subfigure[parameter choices: $\mu=0$, $j=0$, $\chi=\{0.5$ (red), 0.4 (orange), 0.3 (green), 0.2 (cyan), 0.1 (blue)\}]{\includegraphics[width=3.2 in]{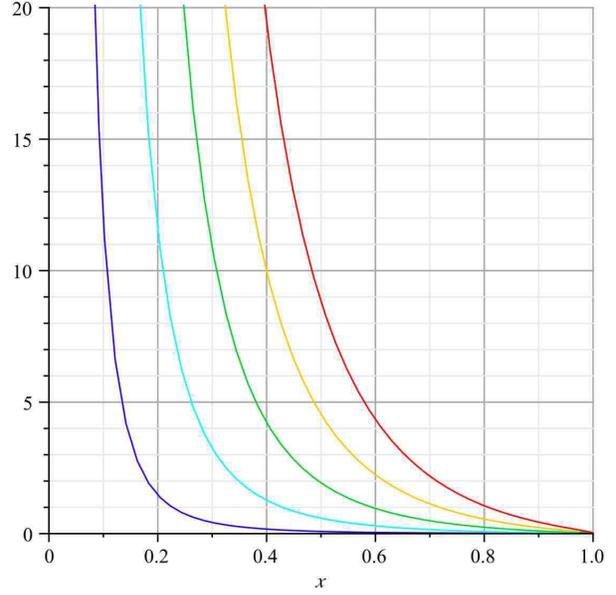}}
     \subfigure[parameter choices: $\mu=1$, $\chi=0.1$, $j=\{2$ (red), 1 (orange), 0 (green), -1 (cyan), -2 (blue)\}]{\includegraphics[width=3.2 in]{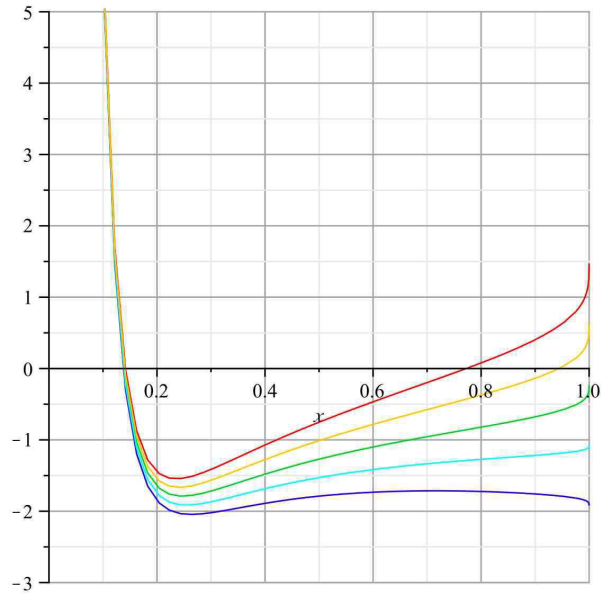}} \\
     \caption{Plots of $\frac{\Psi(r)}{n}$ for the specified choices of parameters, where $j\equiv\ell+\frac{3\chi}{8\pi^2}$; these three plots show the generic behaviour of $\Psi(r)$.  In plots (a) and (c), $\Psi(r)$ is bounded between the red and blue curves as is explained later.  Plot (c) shows the most interesting behavior: because $\Psi(r)$ can go from positive values to negative values at some radius ${\cal R}$; the direction of frame dragging in the space-time is different on each side of ${\cal R}$.}
     \label{fig:P}
\end{figure}

The only thing left to find is $\omega$; using the ansatz (\ref{eq:omega}) we have
\begin{eqnarray}
{\bf d}\omega=\frac{-\Psi'(r)}{f(r)c(r)}{\bf e}^r\wedge {\bf e}^3+\frac{\Psi(r)}{a(r)b(r)}{\bf e}^1\wedge {\bf e}^2\\
\star_4{\bf d}\omega=\frac{-\Psi'(r)}{f(r)c(r)}{\bf e}^1\wedge {\bf e}^2+\frac{\Psi(r)}{a(r)b(r)}{\bf e}^r\wedge {\bf e}^3\nonumber
\end{eqnarray}
In accord with Eq.~(\ref{eq:Gomega}) we conclude that
\begin{equation}
{\bf G^+}=\frac{H^{-1}(r)}{2}\left(\frac{-\Psi'(r)}{f(r)c(r)}+\frac{\Psi(r)}{a(r)b(r)}\right)({\bf e}^r\wedge {\bf e}^3+{\bf e}^1\wedge {\bf e}^2)
\end{equation}
and comparing this equation with (\ref{eq:Gplus}) we find the ordinary differential equation that $\Psi(r)$ must satisfy to be
\begin{equation}
\frac{H^{-1}(r)}{2}\left(\frac{-\Psi'(r)}{f(r)c(r)}+\frac{\Psi(r)}{a(r)b(r)}\right)=\frac{3n\chi}{8\pi^2}\left(\frac{h(r)}{a(r)b(r)}\right)
\end{equation}
which we can manipulate into a more digestible form.  Rearranging, multiplying both sides by $h(r)$ and integrating we find
\begin{equation}
h(r)\Psi(r)=n\ell+\frac{3n\chi}{8\pi^2}\int{H(r)\frac{d}{dr}(h^2(r))dr}
\end{equation}
where $\ell$ is a dimensionless constant of integration. The above integrand can be solved analytically, which is more apparent if we change coordinates from $r$ to $\eta$ and break up the integrand as follows:
\begin{equation}
\Psi(\eta)=\frac{n\ell}{h(\eta)}+\frac{3n\chi}{8\pi^2h(\eta)}\left[\int{(\delta-n^2\mu\eta)\frac{d}{d\eta}(h^2(\eta))d\eta}-2n^2\chi^2\int{\frac{c^2(\eta)}{a(\eta)b(\eta)}h^2(\eta)d\eta}\right]\label{eq:Psi1}
\end{equation}
The first integrand can be integrated by parts (and converted back to $r$) to yield:
\begin{eqnarray}
\int{\left(\delta-n^2\mu\eta\right)\frac{d}{d\eta}(h^2(\eta))d\eta}&=&\left(\delta-n^2\mu\eta\right)h^2(\eta)+n^2\mu\int{h^2(\eta)d\eta}\nonumber\\
&=&\left(\delta-n^2\mu\eta(r)\right)h^2(r)+16\pi^4n^2\mu\left(\eta(r)-\frac{1}{a(r)b(r)}\right)
\end{eqnarray}
where in the second step we have again used the result (\ref{eq:hsq}).  The second integrand of (\ref{eq:Psi1}) is found in \cite{ref:Bena} to have the solution (after converting back to $r$)
\begin{equation}
\int{\frac{c^2(\eta)}{a(\eta)b(\eta)}h^2(\eta)d\eta}=-\frac{8\pi^4}{3}\left(\frac{2c^2(r)}{a^2(r)b^2(r)}-\frac{b(r)c(r)+a(r)c(r)}{a^2(r)b^2(r)}\right)
\end{equation}
so the full solution for $\Psi(r)$ takes the form
\begin{eqnarray}
\Psi(r)&=&\frac{n\ell}{h(r)}+\frac{3n\chi}{8\pi^2}h(r)\left(\delta-n^2\mu\eta(r)\right)+\frac{6\pi^2n^3\chi\mu}{h(r)}\left(\eta(r)-\frac{1}{a(r)b(r)}\right)\nonumber\\
&&+2\pi^2n^3\chi^3\left(\frac{2c^2(r)-c(r)(a(r)+b(r))}{h(r)a^2(r)b^2(r)}\right)
\end{eqnarray}
A plot of $\frac{\Psi(r)}{n}$ for various choices of parameters is given in Fig. \ref{fig:P}

With the equations for $h(r)$, $H(r)$ and $\Psi(r)$ we can find their forms near $r=n\pi$; using the approximations (\ref{eq:approx}) we find
\begin{eqnarray}
h(r)&\approx&\left(\frac{r}{n\pi}-1\right)^{-1/2}+{\cal O}\left(\left(\frac{r}{n\pi}-1\right)^{1/2}\right) \label{eq:hsmall}\\
\eta(r)&\approx&-\frac{\ln 2}{n^2\pi^2}+\frac{1}{2n^2\pi^2}\ln\left(\frac{r}{n\pi}-1\right)+{\cal O}\left(\frac{r}{n\pi}-1\right)\\
H(r)&\approx&\Gamma-\frac{\mu}{2\pi^2}\ln{\left(\frac{r}{n\pi}-1\right)}+\frac{\chi^2}{2\pi^2}\left(\frac{r}{n\pi}-1\right)^{-1}+{\cal O}\left(\left(\frac{r}{n\pi}-1\right)\right)\\
\frac{\Psi(r)}{n}&\approx& \frac{3\chi^3}{2}\left(\frac{r}{n\pi}-1\right)^{-3/2}-\frac{3\mu\chi}{16\pi^4}\left(\frac{r}{n\pi}-1\right)^{-1/2}\ln\left(\frac{r}{n\pi}-1\right)\\
&&+3\chi\left(\frac{\delta}{8\pi^2}+\frac{\mu\ln2}{8\pi^4}-\frac{\chi^2}{4}-\mu\right)\left(\frac{r}{n\pi}-1\right)^{-1/2}+{\cal O}\left(\left(\frac{r}{n\pi}-1\right)^{1/2}\right).\nonumber
\end{eqnarray}
Here $\Gamma=\delta+\frac{\mu\ln{2}}{\pi^2}-\frac{\chi^2}{8\pi^2}$.  Similarly using the approximations (\ref{eq:abclarge}), we find to leading order the large-$r$ approximations to be
\begin{eqnarray}
&&h(r)\approx1+\frac{2n}{r}\label{eq:hlarge}\\
&&\eta(r)\approx\frac{-1}{4n^2}\frac{2n}{r}\\
&&H(r)\approx\delta+\frac{\mu}{4}\frac{2n}{r}+\frac{\chi^2}{4}\left(\frac{2n}{r}\right)^2\\
&&\frac{\Psi(r)}{n}\approx(\ell+\frac{3\delta\chi}{8\pi^2})+\left(\frac{3\delta\chi}{8\pi^2}-\ell-\frac{3\chi\mu}{32\pi^2}(16\pi^4-1)\right)\frac{2n}{r}.
\end{eqnarray} 
We make note of the fact that no matter the values of $\chi$ and $\mu$, the asymptotic value of $H(r)$ is always $\delta$, whose value we can fix by requiring $g_{tt}\rightarrow-1$ as $r\rightarrow\infty$ and hence we have $\delta=1$.

To summarize, in this subsection we have found the differential equations that $h(r)$, $H(r)$ and $\Psi(r)$ must satisfy in order to give a five-dimensional supersymmetric solution on an Atiyah-Hitchin base space.  We have used the exact solutions found in Ref.~\cite{ref:Bena} and we have found Taylor series expansions to these equations in both the  small-$r$ and large-$r$ limits.  A fully explicit analytic solution for arbitrary $r$ is intractable because $\eta(r)$ is defined as an integral of elliptic functions and it is currently unknown whether it has an analytic form.  Fortunately this is unimportant since a large amount of information can be derived from the form of the solution with only qualitative knowledge about the these functions.  The next subsection is devoted to such an approach.

\afterpage{\clearpage}

\subsection{Generic Properties of the Solutions}
\label{sec:Properties}

\subsubsection{The Forms of the Metric}

The full five dimensional metric takes the explicit form
\begin{equation}
ds^2=-\frac{1}{H^2(r)}\left(dt+\Psi(r)\sigma_3\right)^2+H(r)\left(f^2(r)dr^2+a^2(r){\sigma_1}^2+b^2(r){\sigma_2}^2+c^2(r){\sigma_3}^2\right)\label{eq:full}
\end{equation}
which, for convenience can also be written in the expanded form
\begin{equation}
ds^2=-\frac{1}{H^2(r)}dt^2-2\frac{\Psi(r)}{H^2(r)}dt\sigma_3+{\cal G}(r){\sigma_3}^2+H(r)(f^2(r)dr^2+a^2(r){\sigma_1}^2+b^2(r){\sigma_2}^2)\label{eq:expanded}
\end{equation}
where
\begin{equation}
{\cal G}(r)=H(r)c^2(r)-\frac{\Psi^2(r)}{H^2(r)}.
\end{equation}
We also write down the lapse-shift form of the metric:
\begin{equation}
ds^2=-{\cal N}^2(r)dt^2+{\cal G}(r)\left(\sigma_3-\frac{\Psi(r)}{H^2(r){\cal G}(r)}dt\right)^2+H(r)(f^2(r)dr^2+a^2(r){\sigma_1}^2+b^2(r){\sigma_2}^2)\label{eq:lapse}
\end{equation}
where
\begin{equation}
{\cal N}^2(r)=\frac{1}{H^2(r)}\left(1+\frac{\Psi^2(r)}{H^2(r){\cal G}(r)}\right)=\frac{c^2(r)}{H(r){\cal G}(r)}
\end{equation}
is the lapse function.

\subsubsection{Regions with Closed Time-like Curves}

From the metric form of (\ref{eq:expanded}) we can see that something special happens when ${\cal G}(r)$ becomes negative since $g_{\psi\psi}$ also turns negative.  Although $\partial_{\psi}$ is not a Killing vector of the full metric, by examining a congruence of null geodesics we show that the region where ${\cal G}(r)<0$ is one where closed time-like curves (CTCs) are indeed present.  We start by considering the tangent vector to null geodesics
\begin{equation}
k^{\alpha}\partial_{\alpha}=\dot{t}\partial_{t}+\dot{r}\partial_{r}+\dot{\theta}\partial_{\theta}+\dot{\phi}\partial_{\phi}+\dot{\psi}\partial_{\psi}
\end{equation}
where a dot refers to differentiation with respect to the affine parameter, $\lambda$.  We also consider the following four Killing vectors of the full five dimensional metric
\begin{eqnarray}
{\xi_{(t)}}^{\alpha}\partial_{\alpha}=&&\partial_t\nonumber\\
{\xi_{(\phi)}}^{\alpha}\partial_{\alpha}=&&\partial_{\phi}\\
{\xi_{(1)}}^{\alpha}\partial_{\alpha}=&&\sin\phi\partial_{\theta}+\cot\theta\cos\phi\partial_{\phi}-\frac{\cos\phi}{\sin\theta}\partial_{\psi}\nonumber\\
{\xi_{(2)}}^{\alpha}\partial_{\alpha}=&&\cos\phi\partial_{\theta}-\cot\theta\sin\phi\partial_{\phi}+\frac{\sin\phi}{\sin\theta}\partial_{\psi}\nonumber.
\end{eqnarray}
The conserved energy and angular momenta associated with these Killing vectors are
\begin{eqnarray}
-E={\xi_{(t)}}^{\alpha}k_{\alpha},\>\>\>\>\>\>\>\>\>\>\>\>\>
L_{\phi}={\xi_{(\phi)}}^{\alpha}k_{\alpha},\>\>\>\>\>\>\>\>\>\>\>\>\>
L_1={\xi_{(1)}}^{\alpha}k_{\alpha},\>\>\>\>\>\>\>\>\>\>\>\>\>
L_2={\xi_{(2)}}^{\alpha}k_{\alpha}. \label{eq:constraints}
\end{eqnarray}
We wish to study geodesics which are locally non-rotating and hence we choose $L_{\phi}=L_1=L_2=0$ which implies $\dot{\theta}=\dot{\phi}=0$.  We can therefore consider the effective metric on the $(t,r,\psi)$ hypersurface given by
\begin{equation}
d\tilde{s}^2=-\frac{1}{H^2(r)}dt^2-2\frac{\Psi(r)}{H^2(r)}dtd\psi+{\cal G}(r)d\psi^2+H(r)f^2(r)dr^2. \label{eq:effective}
\end{equation}
In this effective metric, $\partial_{\psi}$ is a Killing vector.  When ${\cal G}(r)<0$, we have $\partial_{\psi}$ becoming time-like and it is for this reason that the surface defined by ${\cal G}(r)=0$ is a boundary beyond which CTCs are present.  This boundary, commonly denoted the \emph{velocity of light surface} in the literature, will hereafter be labeled $r_{ctc}$.  Furthermore, we are only concerned with cases where $r_{ctc}\ge n\pi$ so that the velocity of light surface falls within our coordinate range.

The metric (\ref{eq:effective}) can possibly cause some confusion about what it means when ${\cal G}(r)<0$.  If we take $t$=constant slices and $\psi$=constant slices, each time we are left with a time-like hypersurface provided ${\cal G}(r)<0$ so we might be tempted to conclude that there are two time-like directions in this region.  This is nothing more than a coordinate artifact as the determinant of the metric, $det(g)=-f^2(r)c^2(r)$, does not change sign.  Furthermore we can make the coordinate transformation
\begin{equation}
T=t+\Psi(r_{ctc})\psi
\end{equation}
so that the metric takes the form
\begin{equation}
d\tilde{s}^2=-\frac{1}{H^2(r)}(dT+(\Psi(r)-\Psi(r_{ctc}))d\psi)^2+H(r)c^2(r)d\psi^2+H(r)f^2(r)dr^2. 
\end{equation}
In the vicinity of ${\cal G}(r)=0$ (i.e. $r=r_{ctc}$), if we take $T$=constant slices we are left with a space-like metric while taking $\psi$=constant slices yields a time-like one.  This $T$ coordinate gets rid of the confusion caused by the $t$ coordinate but it is inconvenient to work with so we will no longer use it in further analysis.

Utilizing the remaining constraints of (\ref{eq:constraints}), as well as the property that $k^{\alpha}$ is null, we find the tangent vector to our family of null geodesics to be given by
\begin{equation}
{k_{\pm}}^{\alpha}\partial_{\alpha}=E_{\pm}\left(\frac{H(r){\cal G}(r)}{c^2(r)}\partial_{t}\pm \frac{\sqrt{{\cal G}(r)}}{f(r)c(r)}\partial_{r}+\frac{\Psi(r)}{H(r)c^2(r)}\partial_{\psi}\right)\label{eq:k}
\end{equation}
where $+$ and $-$ represent outgoing and ingoing geodesics respectively and $E_{\pm}$ are chosen so that ${k_+}^{\alpha}{k_-}_{\alpha}=-1$.  Upon making the convenient choice $E_-=1$, this normalization implies that at some radius, $r_o$,
\begin{equation}
E_+=\frac{c^2(r_o)}{2H(r_o){\cal G}(r_o)}.
\end{equation}

At the velocity of light surface we have $\frac{dr}{dt}=\pm\frac{c(r)}{f(r)H(r)\sqrt{{\cal G}(r)}}\rightarrow\infty$ and we also have $\frac{dr}{d\lambda}=\pm E_{\pm} \frac{\sqrt{{\cal G}(r)}}{f(r)c(r)}\rightarrow 0$, meaning that null rays cannot cross this surface.  We find the expansion scalar, $\Theta={k^{\alpha}}_{;\alpha}$, of the congruence to be given by
\begin{equation}
\Theta_{\pm}=\pm\left(\frac{{\cal G}'(r)}{c(r)f(r)\sqrt{{\cal G}(r)}}\right)E_{\pm}
\end{equation}
where a prime denotes differentiation with respect to $r$.  In a neighbourhood of $r_{ctc}$, ${\cal G}'(r)>0$ is well behaved and hence the expansions are infinite there.  This shows that the ingoing congruence is converging into a caustic and the outgoing congruence is diverging from a caustic at $r_{ctc}$.  

We can further see that null geodesics are unable to cross the velocity of light surface if we look at the null congruence as a limiting case of a time-like congruence.
Suppose now that instead of a vector field $k^{\alpha}$ tangent to null geodesics we have a vector field $u^{\alpha}$ which is tangent to time-like geodesics.  We still wish to find locally non-rotating solutions so just as before we set $L_{\phi}=L_1=L_2=0$ and we can again consider the effective metric on the $(t,r,\psi)$ plane.  The only difference between $u^{\alpha}$ and our previous $k^{\alpha}$ will be in $\dot{r}$ because of the different normalization conditions for null and time-like tangent vectors.  We find the normalized tangent vector to be
\begin{equation}
{u_{\pm}}^{\alpha}\partial_{\alpha}=E\frac{H(r){\cal G}(r)}{c^2(r)}\partial_t\pm\frac{\sqrt{E^2H(r){\cal G}(r)-c^2(r)}}{c(r)f(r)\sqrt{H(r)}}\partial_r+E\frac{\Psi(r)}{H(r)c^2(r)}\partial_{\psi}
\end{equation}
where $E>0$ is the energy per unit rest mass of the particle on the geodesic.  We can immediately read off $\frac{dr}{d\tau}$ and $\frac{dr}{dt}$, which are given by
\begin{eqnarray}
\frac{dr}{d\tau}=&&\frac{\sqrt{E^2H(r){\cal G}(r)-c^2(r)}}{c(r)f(r)\sqrt{H(r)}}\\
\frac{dr}{dt}=&&\frac{c(r)\sqrt{E^2H(r){\cal G}(r)-c^2(r)}}{Ef(r)H^{3/2}(r){\cal G}(r)}
\end{eqnarray}
and we note that since $H(r)$ is necessarily positive when ${\cal G}(r)$ is positive, we have a radius, $r_{tp}>r_{ctc}$, such that $\sqrt{E^2H(r_{tp}){\cal G}(r_{tp})-c^2(r_{tp})}=0$.  We thus have both $\frac{dr}{d\tau}$ and $\frac{dr}{dt}$ vanishing before the particle reaches $r_{ctc}$, signaling a turning point in the trajectory.  The limit to null geodesics is $E\rightarrow\infty$ and since there is a turning point for all time-like geodesics we see that the velocity of light surface is a turning point for null geodesics.  This merits further discussion.

\begin{figure}
     \centering
     \subfigure[parameter choices: $\mu=3$, $\chi=0.1$, $j=\{ 2$ (red), 1 (orange), 0 (green), -1 (cyan), -2 (blue)\}]{\includegraphics[width=3.2 in]{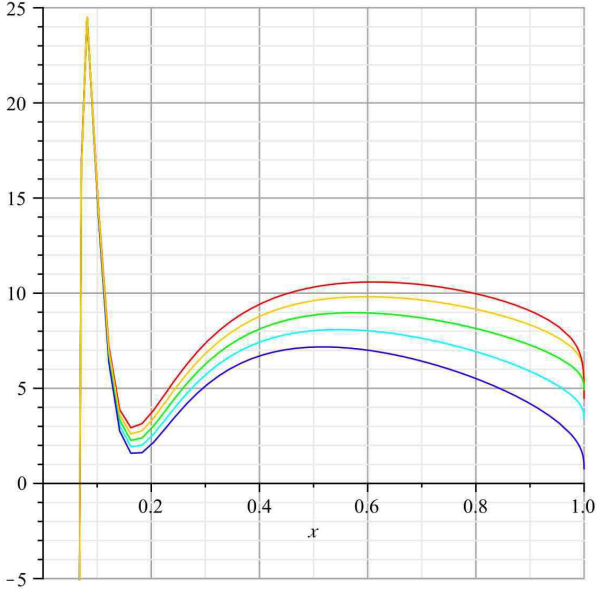}}
     \subfigure[parameter choices: $\mu=4$, $\chi=0.1$, $j=\{ 2$ (red), 1 (orange), 0 (green), -1 (cyan), -2 (blue)\}]{\includegraphics[width=3.2 in]{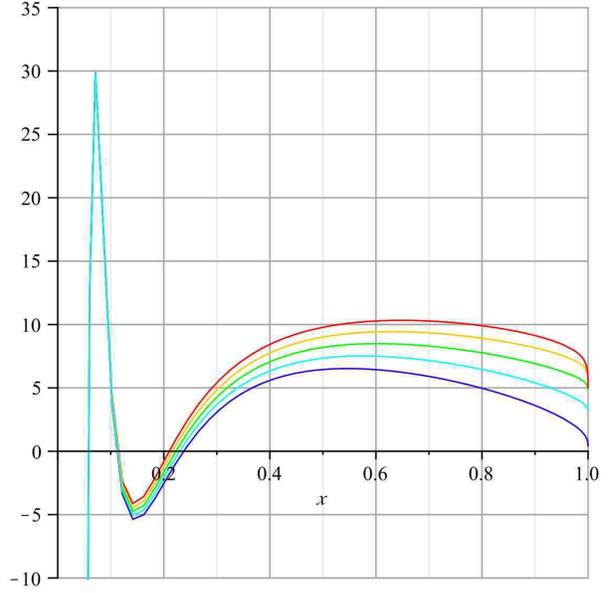}}
     \subfigure[parameter choices: $\mu=2$, $\chi=2/9$, $j=\{ 2$ (red), 1 (orange), 0 (green), -1 (cyan), -2 (blue)\}]{\includegraphics[width=3.2 in]{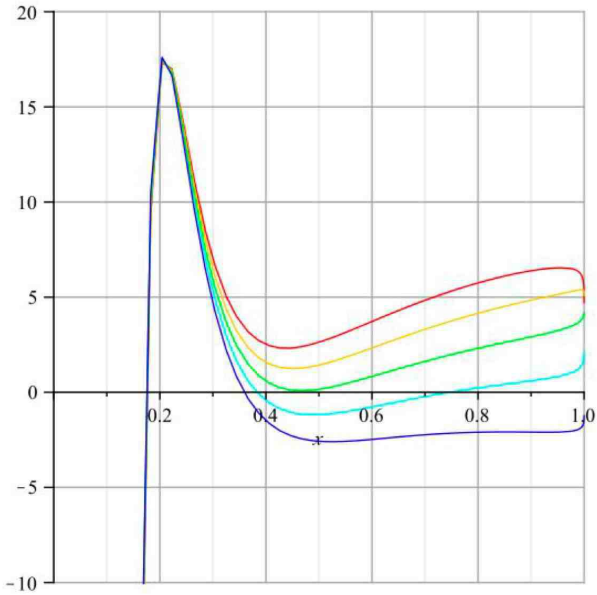}}
     \subfigure[parameter choices: $\mu=0$, $j=0$, $\chi=\{0.5$ (red), 0.4 (orange), 0.3 (green), 0.2 (cyan), 0.1 (blue)\}]{\includegraphics[width=3.2 in]{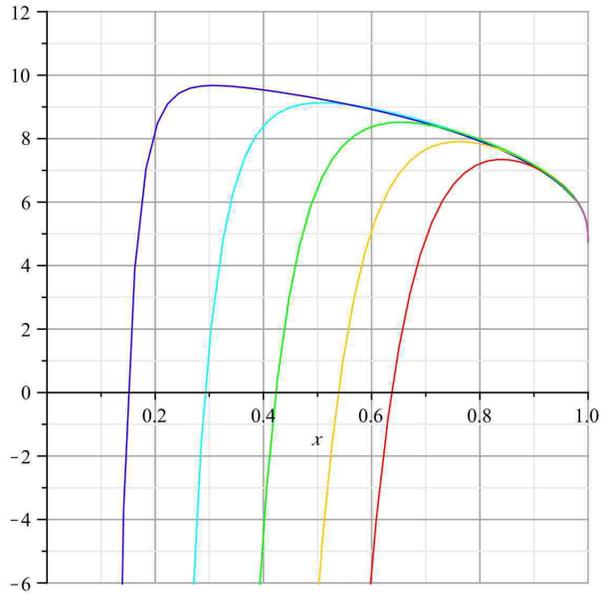}} \\
     \caption{Plots of ${\cal G}(r)$ for the specified choices of parameters, where $j=\ell+\frac{3\chi}{8\pi^2}$.  All four plots demonstrate that there exists at least one radius, $r_{ctc}$, at which ${\cal G}(r)$ changes sign.  Plot (d) shows that if $\mu=j=0$ then there is only one such radius whereas plots (a)-(c) show that there may be multiple such radii in general.  Furthermore, plot (c) shows that there are certain choices of parameters (in this case the green curve) for which both ${\cal G}(r)$ and ${\cal G}'(r)$ vanish at some radius $r_{crit}$.  This surface is similar to the velocity of light surface except that it takes infinite affine parameter to reach.}
     \label{fig:G}
\end{figure}

\afterpage{\clearpage}

This impenetrable velocity of light surface is directly analogous to what appears in the G\"{o}del space-times \cite{ref:Goedel,ref:Boyda,ref:Harmark} and the solutions describing black holes embedded in a G\"{o}del space-time \cite{ref:GimonHashimoto,ref:KernerMann,ref:Behrndt,ref:BehrndtTwo}, hereafter called BH-G\"{o}del solutions.  There are, however, a few very important distinctions that should be pointed out.  The first is that the G\"{o}del solutions are homogeneous, meaning that the existence of CTCs outside of the velocity of light surface implies there exist CTCs through every point in the space-time.  Our solution, on the other hand, is not homogeneous because the Atiyah-Hitchin bolt imposes the notion of a center to the space-time.  This is similar to the broken homogeneity of the BH-G\"{o}del solutions in which the black hole defines the center.  The second distinction is that our solution describes an inverted G\"{o}del-like solution in the sense that the space-time constructed here contains no CTCs for $r>r_{ctc}$, as can clearly be seen from Fig.~\ref{fig:G}.  The G\"{o}del solutions are CTC-free for $r<r_{ctc}$ and the BH-G\"{o}del solutions are CTC-free for $r_H<r<r_{ctc}$ where $r_H$ denotes the horizon of the black hole.  A null ray emanating from the origin in a G\"{o}del space-time (or the horizon in a BH-G\"{o}del space-time) travels out to the velocity of light surface where it forms a caustic and then returns to the origin (or horizon) in finite affine parameter \cite{ref:Goedel,ref:Boyda,ref:GimonHashimoto}.  In our solution the null ray is emitted from infinity, travels inward to the velocity of light surface where it forms a caustic and then returns to infinity.  This process is also done in finite affine parameter as can be seen by integrating $\frac{dr}{d\lambda}= \pm E_{\pm}\frac{\sqrt{{\cal G}(r)}}{f(r)c(r)}$ in the vicinity of $r_{ctc}$.  It is sufficient to show that the null ray can travel from some $r_1$, slightly greater than $r_{ctc}$, to $r_2=r_{ctc}$ in finite affine parameter since for $r>r_{ctc}$ the null ray can usually travel between any two points in finite affine parameter (the only exception is pointed out below).  It should be noted that traveling from infinity to some finite value of $r$ always happens in \emph{infinite} affine parameter but this type of infinity is simply associated with an infinite distance the null rays have to travel and hence is not of interest to us.  We start by Taylor expanding $f(r)$, $c(r)$ and ${\cal G}(r)$ around $r=r_{ctc}$:
\begin{eqnarray}
&&f(r)\approx f(r_{ctc})+f'(r_{ctc})(r-r_{ctc})\\
&&c(r)\approx c(r_{ctc})+c'(r_{ctc})(r-r_{ctc})\\
&&{\cal G}(r)\approx{\cal G}'(r_{ctc})(r-r_{ctc}).
\end{eqnarray}
Next we integrate
$\int_{\lambda_1}^{\lambda_2}d\lambda=\int_{r_1}^{r_{ctc}}\frac{-f(r)c(r)}{\sqrt{{\cal G}(r)}}dr$
and find the solution
\begin{equation}
\Delta\lambda=\frac{2f(r_{ctc})c(r_{ctc})}{\sqrt{{\cal G}'(r_{ctc})}}\sqrt{r_1-r_{ctc}}+{\cal O}((r_1-r_{ctc})^{3/2}).
\end{equation}
This is clearly finite since ${\cal G}'(r_{ctc})\ne0$ as can explicitly be seen in Fig. \ref{fig:G}.  We have thus shown the desired result that null rays travel from large (but finite) $r_o$ down to $r_{ctc}$ and back to $r_o$ in finite affine parameter.  The Atiyah-Hitchin space-time that we have constructed here, then, seems to describe a new type of G\"{o}del-like solution in which the region absent of CTCs includes spatial infinity.  We also note that the norm of the time-like Killing vector, $\partial_t$, can be set to asymptote to unity at infinity so the coordinate time is an appropriate one for observers at infinity.

As a caution, we point out that there are parameter choices for which ${\cal G}(r)=0$ and ${\cal G}'(r)=0$ at some critical radius, $r_{crit}>r_{ctc}$, and so we must check to see whether the same result just derived holds for this surface as well.  The only difference in the analysis is that we replace $r_{ctc}$ by $r_{crit}$ and we also now have
${\cal G}(r)\approx \frac{{\cal G}''(r_{crit})}{2}(r-r_{crit})^2$:
\begin{equation}
\Delta\lambda\approx\frac{\sqrt{2}f(r_{crit})c(r_{crit})}{\sqrt{{\cal G}''(r_{crit})}}\int_{r_1}^{r_{crit}}{\frac{-dr}{r-r_{crit}}}
\end{equation}
which does not converge.  It thus takes an infinite affine parameter for null rays to reach the critical radius.  Furthermore, on these critical surfaces the expansion scalar of the null congruence is finite, so no caustics form and hence these surfaces are not turning points in the trajectory.  Such surfaces causally disconnect the outer region ($r>r_{crit}$) from the inner region ($r<r_{crit}$); as the outer region is no longer geodesically complete   such solutions do not exhibit the same nice behavior as those with just a velocity of light surface.

\subsubsection{Physical Quantities and Parameter Restrictions}

  Using the large$-r$ approximations for the various metric functions, the full metric as $r\rightarrow\infty$ takes the explicit form
\begin{eqnarray}
ds^2\approx&&-\left(1-\frac{\mu}{2}\frac{2n}{r}\right)dt^2+\left(1+\left(\frac{\mu}{4}-1\right)\frac{2n}{r}\right)(dr^2+r^2d{\Omega_2}^2)\label{eq:metricasymptote}\\
&&+n^2\left(4-j^2+\left[\mu+4+j^2\left(4+\frac{\mu}{2}\right)+\frac{3\chi j}{16\pi^2}\left(\mu(16\pi^2-1)-8\right)\right]\frac{2n}{r}\right){\sigma_3}^2 \nonumber\\
&&-2n\left(j+\left[\frac{3\chi}{4\pi^2}-j-\frac{3\chi\mu}{32\pi^2}(16\pi^4-1)\right]\frac{2n}{r}\right)dt\sigma_3.\nonumber
\end{eqnarray} 
where $j\equiv\ell+\frac{3\chi}{8\pi^2}$.  The $dt\sigma_3$ metric element approaches $-nj$ as $r\rightarrow\infty$ so our solution describes a space-time which is rotating at infinity provided $j\ne0$; indeed if we consider radial null geodesics as in Eq.~(\ref{eq:k}) then $\frac{d\psi}{dt}\ne0$ unless $j=0$.  The Riemann tensor vanishes as $r\rightarrow\infty$ and our solution is asymptotically a U(1) fibration  over four dimensional Minkowski space-time; the radius of the circle parameterized by $\psi$ is $n\sqrt{4-j^2}$. 

To calculate the energy and angular momenta, we choose the ``natural" foliation of the space-time by a family of space-like hypersurfaces, $\Sigma_t$, perpendicular to the orbits of ${\xi_{(t)}}^\alpha$.  In this way, it is observers who are locally non-rotating at infinity (so-called ``zero angular momentum" observers) who measure these quantities.  The time-like normal to these hypersurfaces is $n^\alpha=H(r){\xi_{(t)}}^\alpha$ and the radial normal to the boundary, $S_t\equiv\partial\Sigma_t$, of these hypersurfaces is $r_\alpha dx^\alpha=\sqrt{H(r)f^2(r)}dr$.  We note that $\partial_\psi$ is asymptotically a Killing symmetry so we use the Komar formulae:
\begin{eqnarray}
&&E=-\frac{1}{8\pi}\lim_{r\rightarrow\infty}\left[\oint_{S_t(r)}{-2{\xi_{(t)}}^{\beta;\alpha}n_{[\alpha}r_{\beta]}\sqrt{\sigma}d\theta d\phi d\psi}\right]\\
&&J_{\phi,\psi}=\frac{1}{16\pi}\lim_{r\rightarrow\infty}\left[\oint_{S_t(r)}{-2{\xi_{(\phi,\psi)}}^{\beta;\alpha}n_{[\alpha}r_{\beta]}\sqrt{\sigma}d\theta d\phi d\psi}\right]
\end{eqnarray}
where $\sqrt{\sigma}=-\sqrt{H^3(r)}a(r)b(r)c(r)\sin\theta$ is the determinant of the metric induced on $S_t$.   From this, we find the ADM energy and angular momenta to be given by
\begin{eqnarray}
&E=&2n^2\pi\mu, \label{eq:M}\\
&J_{\psi}=& \frac{n^3}{16\pi}\left[16\pi^2j(2-\mu)+3\chi\mu(16\pi^4-1)-24\chi\right],\label{eq:J}\\
&J_{\phi}=& 0.
\end{eqnarray}
The total energy can be positive, negative or zero depending on the value of $\mu$.  Recall that $\mu=\lambda+\chi^2$ so $\lambda$, which very loosely speaking plays the role of the mass of some gravitating object, can contribute positively or negatively to the total energy while $\chi$, which plays the role of a Chern-Simons charge, always contributes positively to the total energy.  The solitonic solutions, which will be shown explicitly shortly, are always found to be solutions with negative total energy.  The fact that $J_{\phi}=0$ is related to our previous discovery that we can consider $\dot{\phi}=0$ for null geodesics; there is no frame dragging around the $\phi$-axis.

We have seen that a velocity of light surface appears whenever ${\cal G}(r)=0$, and now we would like to find solutions that do not possess CTCs at spatial infinity.  Clearly from the metric (\ref{eq:metricasymptote}), the condition we must satisfy is $4-j^2\ge 0$ and thus we have the restriction:
\begin{equation}
-2\le j\le 2
\end{equation}
where equality above leads to another velocity of light surface forming at spatial infinity.  From this inequality and the definition $j=\ell+\chi_0$ (recall $\chi=\frac{8\pi^2}{3}\chi_0$), we interpret $\chi_0$ as being a parameter contributing to a twisting of the space-time, which is not so surprising as its presence is a result of the Chern-Simons term in the action.  When $\chi_0=0$ there is a symmetry in the solution: under $-2\le \ell\le2$ the sense of rotation depends on the sign of $\ell$.  When $\chi_0\ne 0$ this symmetry is destroyed because of the extra rotation that $\chi_0$ supplies; instead the symmetry is in the form $-2<\ell+\chi_0<2$.  If we consider $\chi_0>0$ then the restriction on $\ell$ is $-(2+\chi_0)\le \ell\le (2-\chi_0)$ but if we consider sending $\chi_0\rightarrow -\chi_0$ then the restriction becomes $-(2-\chi_0)\le \ell\le(2+\chi_0)$.  Changing the sign of $\chi_0$ thus changes the sign of the restriction on $\ell$; the sign of $\chi_0$ determines the handedness of the extra rotation supplied by $\chi_0$.

\subsubsection{Singularities and Solitons}

We can use the small$-r$ approximations along with the rotated axes of (\ref{eq:rotate}) to write down the asymptotic form of the metric near $r=n\pi$.  If $\chi\ne0$ then the metric goes to
\begin{eqnarray}
ds^2&\approx& -\frac{4\pi^4}{\chi^4}\left(\frac{r}{n\pi}-1\right)^2dt^2-\frac{12n\pi^4}{\chi^2}\left(\frac{r}{n\pi}-1\right)^{1/2}dt\tilde{\sigma}_2-9n^2\pi^4\chi^2\left(\frac{r}{n\pi}-1\right)^{-1}{\tilde{\sigma}_2}^2\nonumber\\
&&+\frac{\chi^2}{2\pi^2}\left(\frac{r}{n\pi}-1\right)^{-1}\left(dr^2+4(n\pi)^2\left(\frac{r}{n\pi}-1\right)^2{\tilde{\sigma}_3}^2+(n\pi)^2({\tilde{\sigma}_1}^2+{\tilde{\sigma}_2}^2)\right)
\end{eqnarray}

Note that we no longer have the SO(3) orbit collapsing in dimensionality on the bolt, which can be seen from the determinant of the metric:
\begin{equation}
det(g)=-H^2(r)f^2(r)a^2(r)b^2(r)c^2(r)\sin^2\theta.
\end{equation}
Near $r=n\pi$ the determinant approaches a constant value since $H(r)\sim\left(\frac{r}{n\pi}-1\right)^{-1}$ and $a(r)\sim\left(\frac{r}{n\pi}-1\right)$ while all of the other functions approach constant values.  The ``bolt" is still a singularity because both the Kretschmann and Ricci scalars diverge there.  For instance, for $\chi\ne0$ we find the Ricci scalar to leading order to be
\begin{equation}
R\approx\frac{-3}{n^2\chi^2}(15\pi^4-1)\left(\frac{r}{n\pi}-1\right)^{-1},
\end{equation}
for $\chi=0$, $\mu\ne0$ it is given by
\begin{equation}
R\approx\frac{1}{\mu n^2\pi}\left(\left(\frac{r}{n\pi}-1\right)^2\ln\left(\frac{r}{n\pi}-1\right)^3\right)^{-1},
\end{equation}
and for $\chi=\mu=0$ it reduces to
\begin{equation}
R\approx\frac{2}{(n\pi)^2}+\frac{j^2}{4n^2\pi^4}\left(\frac{r}{n\pi}-1\right)^{-1}
\end{equation}
We thus have a singularity at $r=n\pi$ for all choices of parameters except the very specific choice $\chi=\mu=j=0$.  We were unable to find any solutions that exhibit an event horizon so the curvature singularity at $r=n\pi$ is not hidden behind such a surface.  However, if a velocity of light surface is present then the singularity is always hidden behind it and so is not truly naked.  

The singularity at $r=n\pi$ is not the only singularity potentially present in the space-time; from the form of the metric given in Eq.~(\ref{eq:full}), it is easy to see that if $H(r)$ is negative then the space-time attains a Euclidean signature $(-----)$ meaning that at some radius, $r_s>n\pi$, such that $H(r_s)=0$, there is a solitonic boundary.  The solution generating technique outlined in section 2 is applicable as long as $H(r)\ge0$ and globally defined, so we are forced to constrict the range of our radial coordinate to some region where $H(r)\ge0$, in this case $[r_s,\infty)$.  If there are two radii, $r_{s1}$ and $r_{s2}$ such that $r_{s1}<r_{s2}$, at which $H(r)=0$, the space-time is cut off at the outermost radius, $r_{s2}$.  $H(r)$ for the Atiyah-Hitchin solution is indeed easy to make negative for some $r_s>n\pi$; in examining Fig.~\ref{fig:H} we can see that we can only get negative values of $H(r)$ if we take $\mu<0$; if $\chi=0$ any negative value of $\mu$ will always yield a solitonic surface but if $\chi\ne0$ then $\mu$ must be sufficiently large and negative to produce such a surface.  The surface of the soliton is always singular, which can be seen from the Ricci scalar expanded in inverse powers of $H(r)$:
\begin{equation}
R=\left(\frac{(\Psi^2(r))'}{2f(r)a(r)b(r)c(r)}\right)\frac{1}{H^{4}(r)}-\left(\frac{(H'(r))^2}{2f^2(r)}\right)\frac{1}{H^{3}(r)}+{\cal O}\left(\frac{1}{H^2(r)}\right)+{\cal O}\left(\frac{1}{H(r)}\right)
\end{equation}
where the ${\cal O}(H^{-2}(r))$ and ${\cal O}(H^{-1}(r))$ terms are too lengthy to write down here and for our purposes are not overly interesting.  Near $r=r_s\ne n\pi$, $H(r)\approx H'(r_s)(r-r_s)$ while all the other metric functions are non-zero.  If we make a generic choice of parameters, then $\Psi(r)\ne 0$ and the curvature diverges like $(r-r_s)^{-4}$.  For the specific choice of $\chi=j=0$ and $\mu<0$, we have $\Psi(r)=0$ and $H'(r)\ne0$; in this case the curvature diverges like $(r-r_s)^{-3}$, which is the least singular behaviour we can obtain.  For this choice of parameters there is no rotation, meaning there is no velocity of light surface and hence the singularity is truly naked.  

We are again unable to find any event horizons present in the space-time but for generic choices of parameters we have the singularity at $H(r)=0$ shielded by the velocity of light surface.  This is because at the boundary when $H(r)\rightarrow 0$, we must have one of three cases: $(i)$ ${\cal G}(r)\rightarrow-\infty$ because of the $-H^{-2}(r)$ dependence, $(ii)$ ${\cal G}(r)<0$ if $\Psi(r)\propto H(r)$ or $(iii)$ ${\cal G}(r)=0$ if we can arrange $\Psi(r)\propto H^p(r)$ where $p>1$.  The latter is impossible to arrange because of the form of the solution for $\Psi(r)$ along with the behavior of the metric functions $a(r)$, $b(r)$ and $c(r)$.  We conclude, then, that when $H(r)=0$, ${\cal G}(r)<0$ and hence the velocity of light surface is necessarily outside of the solitonic boundary.  The only exception to this is for $\chi=j=0$ in which case $\Psi(r)=0$ and there is no velocity of light surface and hence the singularity is not shielded.

To complete the present discussion we note that the Ricci and Kretschmann curvature invariants are perfectly regular on the velocity of light surface, as is the field strength; within a small vicinity of the velocity of light surface the space-time and gauge field are smooth and well behaved.  This surface, just as in the G\"{o}del solutions, is not a physical boundary to the space-time but for all practical purposes it acts like one since the interior is completely causally disconnected from the exterior.  Because of this causal disconnectedness from the interior along with the exterior space-time being geodesically complete, one is free to regard the exterior of the velocity of light surface as comprising the whole space-time.

\section{Discussion}
\label{sec:Conclusion}

We have constructed a solution to five dimensional minimal supergravity using an Atiyah-Hitchin base space.  The Atiyah-Hitchin metric does not admit a triholomorphic Killing vector field and hence there is no Gibbons-Hawking form of the metric to be exploited.  Instead we employed the semi-analytic solutions found in \cite{ref:Bena} and were able to solve these equations to leading order near the Atiyah-Hitchin bolt and at asymptotic infinity.  While a fully explicit analytic solution does not exist for arbitrary choices of the radial coordinate, we were able to easily perform the necessary numerical integration to explicitly show the general form of the solutions. 

By considering the general form of the space-time metric and using arguments based on the structures of the metric functions we were able to show that our solution describes space-times in which there are singular surfaces present.  Such singular surfaces are either the original bolt from the Atiyah-Hitchin base space or a solitonic boundary where the signature of the space-time turns Euclidean.  We have found no event horizons, so the singular surfaces are naked singularities in the usual sense.  The solutions were also shown to typically include a region of CTCs where the effective Killing vector $\partial_{\psi}$ turns time-like.  The boundary of this region was found to always be at a greater radius than the singular surface and null rays are unable to cross this boundary so the naked singularities are found to be typically masked by this velocity of light surface.  There is only a small subset of parameter space, including $\chi=j=0$ and $\mu<0$, in which there is no region of CTCs and for which the singularity is not masked but truly naked.

In our solution we have made the very specific choice of taking the scalar function $H$ to be a function of radial coordinate only.  This is because such a choice is the easiest case to consider; there surely are further solutions corresponding to differing choices of $H$.  Because of the equation that $H$ must satisfy, namely $\Delta H(x^a)=\frac{4}{9}({\bf G^+})^2$, unless ${\bf G^+}=0$ we would not be able to make $H$ a separable function of $x^a$ and a completely general solution for $H$ would be extremely difficult to find.  If ${\bf G^+}=0$, however, then $H$ can be made separable and at least asymptotically a solution can be found.  We predict that because $a(r)$, $b(r)$ and $c(r)$ are nonzero everywhere except when $a(r)=0$ at $r=n\pi$, if the Atiyah-Hitchin base space does admit solutions with black holes then the potential black holes will necessarily be extremal and $H$ will most likely be a function of both radial and angular coordinates.  All of these issues are
  currently left for future work.

 It is well known that quantum particles are able to tunnel through classical barriers.  In the context of gravitating bodies, this phenomenon gives rise to thermodynamic descriptions of black holes based on the emission spectra of such particles.  It would be an interesting problem to analyze whether it is possible in the space-time constructed in this paper for quantum particles to tunnel between the classically causally disconnected regions separated by the velocity of light surface.  If such a phenomenon is possible then the study of this space-time from a thermodynamic perspective could be explored.  With such a thermodynamic description, it would be a further interesting problem to describe such thermodynamics in terms of a microstate counting.  We leave these issues for further consideration.

The solution constructed in this paper bears some similarity to the G\"{o}del  and BH-G\"{o}del solutions previously constructed in \cite{ref:Gauntlett,ref:Goedel,ref:Boyda,ref:Harmark,ref:GimonHashimoto,ref:KernerMann,ref:Behrndt,ref:BehrndtTwo}.   One key difference is that the G\"{o}del space-time is homogeneous and so has CTCs through every point whereas our solution does not have this property because of the existence of the bolt at $r=n\pi$.  One rather striking and interesting feature of the solution constructed herein is that unlike the (BH-)G\"{o}del solutions the region containing closed time-like curves is contained entirely within the bulk of the space-time and spatial infinity is seemingly free of pathologies.  It is conjectured in Ref.~\cite{ref:Behrndt} that whenever CTCs develop in the bulk, the dual CFT is pathological and not well-defined. The idea is that the CFT metric itself develops CTCs and we thus would not be able to make sense of a quantum field theory on a space-time with CTCs.  This argument, however, presumes space-times that possess CTCs at asymptotic infinity so it is not surprising that the CFT should suffer pathologies.  The space-time constructed in this paper, however, does not suffer from CTCs at infinity so it would be worthwhile to see if the same argument holds here. If there is a holographic interpretation of our solution, it would be an interesting counter-example to study if the dual CFT were free of pathologies.

\bigskip
\section*{Acknowledgements}
This work was supported by the Natural Sciences and Engineering Research Council of Canada.

\bibliographystyle{plain}

\end{document}